\documentstyle[prd,aps]{revtex}

\newcommand{\bmit}[1]{\mbox{\boldmath $#1$}}

\begin{document}
\draft
\preprint{Accepted for publication in \prd}
\title{Post--Newtonian Lagrange Planetary Equations}
\author{Mirco Calura, Pierluigi Fortini and Enrico Montanari}
\address{Department of Physics, University of Ferrara and
INFN Sezione di Ferrara, Via Paradiso 12, I-44100 Ferrara, Italy}
\address{calura@axpfe1.fe.infn.it\\
fortini@axpfe1.fe.infn.it\\
montanari@axpfe1.fe.infn.it}
\maketitle
\begin{abstract}
We present a method to study the time variation of 
the orbital parameters of a Post--Keplerian binary system undergoing 
a generic external perturbation. The method is the 
relativistic extension of the planetary Lagrangian equations.  
The theory only assumes the smallness of the external perturbation 
while relativistic effects are already included in the unperturbed 
problem. This is the major advantage of this novel approach over 
classical Lagrangian methods.
\end{abstract}



\section{Introduction}
Since the early stages of classical celestial mechanics, a 
very large amount 
of efforts has been made in order to find exact or approximate 
solutions to the problem of N point--like interacting bodies. 
It is well known that a general solution exists only for $N=2$ 
(Keplerian solution). In general, when N exceeds two searching for 
even approximate solutions becomes a very difficult task.
Fortunately many N--body systems of 
interest in celestial mechanics may be considered as a 2--body 
problem, and the interaction of the other $N-2$ bodies can be regarded 
as a perturbation.
In these cases the distance of the $N-2$ bodies, together with the 
relative magnitude of masses, allows one to resort to standard 
perturbative methods (see for instance \cite[ch.6]{mac}).
Moreover when a 2--body system suffers perturbations such as drag or 
radiation damping forces, oblateness of one of the two bodies and so 
on, the usual Keplerian solution must be viewed only as a 
zeroth--order approximation.

In the framework of newtonian gravity, all these kinds of 
perturbations 
may be handled using the Lagrangian planetary equations through
which the time dependence of orbital elements 
(otherwise constant) is achieved. 
This way the motion of the system is formally  
Keplerian but the orbital elements are allowed to 
vary with time. This procedure has been widely used from 
physicists and astronomers to study 
Newtonian binary systems; but what happens if relativistic effects 
are to be taken into account together with the external 
perturbation? 
A first way to approach the problem is to assume that
relativistic effects and the external perturbation are roughly of 
the same order of magnitude; in this {\em semiclassical} standpoint 
they are both considered as a perturbation to Keplerian motion.

We propose a novel procedure to account for relativistic effects in 
the unperturbed problem, in which the only perturbation is the 
external one. For this scheme we divide our work in two steps; first 
the relativistic two--body problem needs to be solved (this has 
already been done up to the 
1PN and 2PN orders beyond the classical limit by various authors 
(see, for instance, \cite{1pn} and \cite{2pn}); this is the 
so--called post--keplerian solution); then our task, that is the 
main result of this paper, is to find out the 
relativistic version of Lagrangian planetary equations,
giving the time dependence of relativistic orbital elements.

\section{Post--Keplerian solution}
Let us consider a system of two bodies, moving under mutual 
gravitational attraction. Since we want to focus our attention on  
systems whose relativistic effects are not negligible, 
we have to introduce a Post--Newtonian parameter
quantifying the relevance of these effects; this 
parameter is defined as follows
\begin{equation}
\epsilon_{PN}\stackrel{def}= \frac{v}{c}
\label{parpost}
\end{equation} 
In the above expression $v$ is the typical speed of the 
bodies of the system and $c$ is the speed of light; 
using relativistic Einstein equations 
it can be shown that the equations of motion up to 
the 1PN order, i.e. retaining only the terms in 
$(\epsilon_{PN})^2$, may be derived from the following Lagrangian,
in a reference frame with the 1PN centre of mass at rest
(see for instance~\cite{1pn}):
\begin{eqnarray}
&&{\cal L}_0 = \frac{1}{2}\, \frac{\bmit{v}^2}{c^2} + 
\frac{G\,m}{c^2\,r} + \frac{1}{8}\, 
\left( 1 - 3\nu\right)\, \frac{\bmit{v}^4}{c^4} + 
\frac{G\,m}{2c^2\, r}\, \left(
\left(3 + \nu\right)\, \frac{\bmit{v}^2}{c^2} + \nu\, \frac{
(\bmit{v}\cdot\bmit{x})^2}{c^2\,r^2}\right) - 
\frac{G^2\,m^2}{2c^4\, r^2}\nonumber\\
\label{L0}
\end{eqnarray}
where
\begin{eqnarray}
&&\bmit{x} = \bmit{r_2} - \bmit{r_1}\nonumber\\
&&\bmit{r_B} = \frac{{\cal E}_1}{{\cal E}_1+{\cal E}_2}\, \bmit{r_1} + 
\frac{{\cal E}_2}{{\cal E}_1+{\cal E}_2}\, \bmit{r_2}\nonumber\\
&&{\cal E}_1 = m_1 c^2 + \frac{m_1}{2}\,\bmit{v_1}^2 - 
\frac{G m_1m_2}{2r}
\nonumber\\
&&{\cal E}_2 = m_2 c^2 + \frac{m_2}{2}\,\bmit{v_2}^2 - 
\frac{G m_1m_2}{2r}
\nonumber\\
&&m = m_1 + m_2,\qquad \nu = \frac{m_1 m_2}{m^2}\nonumber\\
&&r = |\bmit{x}|,\qquad \bmit{v} = \frac{d\bmit{x}}{dt}\nonumber
\end{eqnarray}
In the above formulae, $\bmit{r_1}$ and $\bmit{r_2}$ are the radius 
vectors of the two bodies, $m_1$ and $m_2$ denote their 
masses and $\bmit{r_B}$ is the position of the Post--Newtonian 
centre of mass (see \cite[eqs.(5.4.5)--(5.4.9)]{PNlibro}).   
In the following it will become clear the usefulness of the 
Hamiltonian approach with respect to the Lagrangian one; therefore 
we calculate the canonical momenta deriving from (\ref{L0}):
\begin{eqnarray}
\bmit{p} = \frac{\partial {\cal L}_0}{\partial \bmit{v}} = 
\frac{\bmit{v}}{c^2} + \frac{1}{2}\, \left(1 - 3\nu\right)\, 
\frac{\bmit{v}^2\, \bmit{v}}{c^4} + \frac{G\,m}{c^2\,r}\, 
\left(3 + \nu\right)\, 
\bmit{v} + \frac{G\,\nu m}{c^2}\, \frac{(\bmit{v}\cdot\bmit{x})}{r^3}\, 
\bmit{x}
\label{p}
\end{eqnarray}
The Hamiltonian function of the system is defined as usual: 
\begin{equation}
{\cal H}_0 = \bmit{p}\cdot\bmit{v} - {\cal L}_0
\end{equation}
Now, from (\ref{L0}) and (\ref{p}) we are in a position to 
express explicitly such an Hamiltonian in terms of the canonical 
variables $\bmit{x}$ and $\bmit{p}$, obtaining:
\begin{eqnarray}
&&{\cal H}_{0} = \frac{c^2}{2}\,\bmit{p}^2 - \frac{G\,m}{c^2\,r} - 
\frac{c^4}{8}\,
\left(1 - 3 \nu\right)\, \bmit{p}^4 - \frac{G\,m}{2 r}\,
\left(3 + \nu\right)\, \bmit{p}^2 - \frac{G\,m \nu}{2 r^3}\,
\left(\bmit{x}\cdot\bmit{p}\right)^2 + \frac{G^2\,m^2}{2c^4\,r^2}\nonumber\\
&&
\label{H0}
\end{eqnarray}
Hence, Hamilton equations are:
\begin{eqnarray}
&&\frac{d\bmit{p}}{dt} = -\,\frac{G\,m}{c^2\,r^3}\,\bmit{x} - 
\frac{G\,m}{2 r^3}\, 
\left(3 + \nu\right)\,\bmit{p}^2 \bmit{x} - \frac{3 G\, m \nu}{2 r^5}\,
\left(\bmit{x}\cdot\bmit{p}\right)^2\,\bmit{x} + \frac{G\,m \nu}{r^3}\,
\left(\bmit{x}\cdot\bmit{p}\right)\,\bmit{p} + 
\frac{G^2\,m^2}{c^4\,r^4}\,\bmit{x}\nonumber\\
&&\label{eqmoto0p}\\
&&\frac{d\bmit{x}}{dt} = c^2 \bmit{p} - \frac{c^4}{2}\,\left(1 - 3\nu\right)
\,\bmit{p}^2 \bmit{p} - \frac{G\,m}{r}\,\left(3 + \nu\right)\,\bmit{p} - 
\frac{G\,m \nu}{r^3}\,\left(\bmit{x}\cdot\bmit{p}\right)\,\bmit{x}
\label{eqmoto0x}
\end{eqnarray}
The exact solution (called Post--Keplerian), up to 1PN order, has been 
found by Damour and Deruelle (see \cite{1pn}, and also \cite{2pn} for 
its extension to 2PN level). 
The Post--Keplerian solution may be written in a form very similar 
with respect to the classical one as follows: 
\begin{eqnarray}
&&x^1 = \left(\cos{\omega}\, \cos{\Omega} - 
\sin{\omega}\, \sin{\Omega}\, \cos{i}\right)\, \tilde{x}^1 -
\left(\sin{\omega}\, \cos{\Omega} + 
\cos{\omega}\, \sin{\Omega}\, \cos{i}\right)\, \tilde{x}^2
\nonumber\\ 
&&x^2 = \left(\cos{\omega}\, \sin{\Omega} + 
\sin{\omega}\, \cos{\Omega}\, \cos{i}\right)\, \tilde{x}^1 -
\left(\sin{\omega}\, \sin{\Omega} - 
\cos{\omega}\, \cos{\Omega}\, \cos{i}\right)\, \tilde{x}^2
\nonumber\\
&&x^3 = \sin{\omega}\, \sin{i}\  \tilde{x}^1 + 
\cos{\omega}\, \sin{i}\  \tilde{x}^2 
\label{eulero}
\end{eqnarray}

where

\begin{eqnarray}
&&\tilde{x}^1 = r\, \cos\left(2\, \left(1 + \kappa\right)\, 
\arctan\left(\sqrt{\frac{1 + e_3}{1 - e_3}}\, \tan\left(
\frac{\eta}{2}\right)\right)\right) = r\,\cos{\tilde\eta}
\nonumber\\
\label{xtilde}\\
&&\tilde{x}^2 = r\, \sin\left(2\, \left(1 + \kappa\right)\,
\arctan\left(\sqrt{\frac{1 + e_3}{1 - e_3}}\, \tan\left(
\frac{\eta}{2}\right)\right)\right) = r\,\sin{\tilde\eta}
\nonumber
\end{eqnarray}

and

\begin{eqnarray}
&&n\, \left(t - T\right) = \eta - e_1\, \sin{\eta}\label{Teta}\\
&&r = a\, \left(1 -e_2\, \cos{\eta}\right)\\
&&n = \sqrt{\frac{G\,m}{a^3}}\, \left(1 - \frac{(9 -\nu)}{2}\, 
\frac{G\,m}{c^2\,a}\right)\\
&&\kappa = \frac{3 G\, m}{c^2\,a (1-e^2)}\label{kappa}\\
&&e_1 = \left(1 - \frac{(8 - 3\nu)}{2}\, \frac{G\,m}{c^2\,a}\right)\, e\\
&&e_2 = e\\
&&e_3 = \left(1 + \frac{\nu}{2}\, \frac{G\,m}{c^2\,a}\right)\, e
\label{e3}
\end{eqnarray}

In these equations $\omega$, $\Omega$ and $i$ 
are Euler angles, defining the rotation that connects the 
observation reference frame with the intrinsic frame of the 
motion. In celestial mechanics they are usually referred to as 
{\em argument of periastron} (the angle in orbital plane 
from the line of nodes 
(see~\cite[\S 4.4]{gold}) to the perihelion point), {\em longitude of 
the ascending node} (the angle measured from the positive $x$ axis of 
the observer to the line of nodes) and {\em inclination of the orbit} 
(the angle between the orbital plane and the $x$--$y$ plane of the 
observer), respectively.
The other elements of the orbit are the {\em semimajor axis} of the 
ellipse $a$, the {\em eccentricity} $e$ and the {\em 
time of periastron} passage $T$.

\section{1PN Lagrangian brackets}
In last section we have reviewed the post--newtonian solution of a 
binary system of bodies. Such a solution is quite correct 
provided that either the system is completely isolated or the 
external perturbation induces effects whose order of magnitude is less 
than or equal to the 2PN ones; if we quantify the weakness of 
external perturbation to the parameter $\epsilon_{ext}$, our 
last assumption is:
\begin{equation}
\epsilon_{ext} \leq (\epsilon_{PN})^2
\end{equation}
In the framework of the validity of the above expression the simple 
post--Keplerian solution is correct. When the external perturbation 
is so strong that the above expression does not hold true anymore, 
we have to take into 
account its effect on the 1PN relative motion of the two bodies. 
To this purpose we have developed a perturbation method enabling us to 
calculate the time dependence of orbital parameters due to external 
perturbation. 
Such a method may be viewed as the relativistic extension of the 
planetary Lagrangian equations, which are widely being used in 
classical celestial 
mechanics to evaluate the time evolution of the orbital elements for 
Keplerian motion. Indeed it is our aim to calculate 
the time evolution of orbital parameters for Post--Keplerian motion.
To summarize we assume that the solution of the perturbed problem has 
the same functional dependence on orbital 
parameters and on time explicitly as the unperturbed problem.

In the presence of an external perturbation,
the Hamiltonian describing the system  can be written as:
\begin{equation}
{\cal H} = {\cal H}_0(\bmit{x},\, \bmit{p}) + 
{\cal H}_1(\bmit{x},\, \bmit{p})
\label{Hslpitting}
\end{equation}
where ${\cal H}_0$ is the usual 2--body Hamiltonian, which is exact 
up to 1PN order, while ${\cal H}_1$ describes the most 
general perturbation. This latter is a sort of {\em 
disturbing function} with the only difference that the usual 
classical disturbing function (see \cite[chs.5, 6]{mac} and 
\cite{browclem} if the external perturbation is due to 
the gravitational attraction of other bodies) is a perturbation to the
{\em Lagrangian}, while the one concerned through this paper is  a 
perturbation to the {\em Hamiltonian}. With the above assumptions, the 
solution fulfills the following equations:
\begin{equation}
\frac{\partial\bmit{x}}{\partial t} = 
\frac{\partial {\cal H}_0}{\partial\bmit p}\ ,\qquad\qquad\qquad
\frac{\partial\bmit{p}}{\partial t} =
- \frac{\partial {\cal H}_0}{\partial\bmit x}
\label{eqimp}
\end{equation}
where $\bmit{x} = \bmit{x}(a,e,T,\omega,\Omega,i,t)$ and 
$\bmit{p} = \bmit{p}(a,e,T,\omega,\Omega,i,t)$.

As in the classical case, we need to find out 
the equations governing the time dependence of 
orbital parameters ($a,\,e,\,T,\,\omega,\,\Omega,\,i$) due to an
external perturbation. First of all we have to calculate the 
relativistic Lagrangian brackets defined as 
follows~\cite[\S 9--4]{gold}:  
\begin{equation}
\left[C_j,\, C_k\right] \stackrel{def}= 
\frac{\partial\bmit{x}}{\partial C_j}\cdot
\frac{\partial\bmit{p}}{\partial C_k} - 
\frac{\partial\bmit{x}}{\partial C_k}\cdot
\frac{\partial\bmit{p}}{\partial C_j}
\label{defparentesi}
\end{equation} 
where $C_j$ with $1 \le j \le 6$ are the orbital elements 
$a,\,e,\,T,\,\omega,\,\Omega,\,i$. On the strength of
equations~(\ref{eqimp}) it is easy to show that the Lagrangian 
brackets do not depend explicitly upon time (for the classical version 
of this property see~\cite[\S 6--2]{mac}). 

From definition~(\ref{defparentesi}) and 
Hamilton equations of motions pertaining eq.~(\ref{Hslpitting}), 
we find, after straightforward calculation (see for 
instance~\cite[\S 11--2]{gold}): 
\begin{equation}
\sum_{k = 1}^{k = 6}\left[C_j,\, C_k\right]\, \frac{d C_k}{dt} = 
-\, \frac{\partial{\cal H}_1}{\partial C_j}
\label{sistema}
\end{equation}
We remark that this latter equations are rather similar  
to the classical ones (see \cite[p.133, eq.(6--19)]{mac}), 
where ${\cal H}_1$ plays the role of 
disturbing function (the sign is reversed since our 1PN disturbing 
function is ascribed to an Hamiltonian).

Since 1PN Lagrangian brackets do not depend explicitly upon time, it 
suffices to evaluate them at the periastron (i.e. at $t = T$). 
In doing so we follow 
the same procedure as in the classical case (see for instance 
\cite[\S 6--3]{mac}), 
i.e. we expand in powers of $t - T$, up to second--order terms, 
the relative radius vector $\bmit{x}$ and the canonical momentum 
$\bmit{p}$, then we compute Lagrangian 
brackets~(\ref{defparentesi}); finally we set $t = T$. 
After very long but straightforward calculation we achieve the only 
non--vanishing brackets: 
\begin{eqnarray}
&&\left[a,\, T\right] = \frac{G\,m}{2 c^2\,a^2}\, \left(1 + 
\frac{G\,m}{2 c^2\, a}\, \left(\nu-7\right)\right)
\label{aT}\\
\nonumber\\
&&\left[a,\, \omega\right] = -\frac{a \sqrt{1 - e^2}}{2 c^2}\, 
\sqrt{\frac{G\,m}{a^3}}\, \left(1 -\frac{2 G\,m}{c^2\,a (1 - e^2)}\, \left(
1 + \frac{e^2}{2} - \frac{\nu e^2}{4}\right)\right)\\
\nonumber\\
&&\left[a,\, \Omega\right] = 
-\frac{a \sqrt{1 - e^2}}{2 c^2}\, \sqrt{\frac{G\,m}{a^3}}\, \left(1 - 
\frac{2 G\,m}{c^2\,a (1-e^2)}\, \left(1 + \frac{e^2}{2} - 
\frac{\nu e^2}{4}\right)\right)\, \cos{i}\\
\nonumber\\
&&\left[e,\, \omega\right] = 
\frac{a^2 e}{c^2\,\sqrt{1 - e ^2}}\, \sqrt{\frac{G\,m}{a^3}}\, \left(
1 - \frac{4 G\,m}{c^2\,a (1 - e^2)}\, \left(1 - \frac{e^2}{4} - 
\frac{\nu}{4} + \frac{\nu e^2}{8}\right)\right)\\
\nonumber\\
&&\left[e,\, \Omega\right] = 
\frac{a^2 e}{c^2\,\sqrt{1 - e ^2}}\, \sqrt{\frac{G\,m}{a^3}}\, \left(
1 - \frac{4 G\,m}{c^2\,a (1 - e^2)}\, \left(1 - \frac{e^2}{4} -
\frac{\nu}{4} + \frac{\nu e^2}{8}\right)\right)\, \cos{i}\nonumber\\
\\
&&\left[\Omega,\, i\right] = 
- \frac{a^2}{c^2} \sqrt{1 - e ^2}\, \sqrt{\frac{G\,m}{a^3}}\, \left(1 + 
\frac{2 G\,m}{c^2\,a (1 - e^2)}\, \left(1 + \frac{e^2}{2} - 
\frac{\nu e^2}{4}\right)\right)\, \sin{i}
\end{eqnarray}

\section{1PN planetary Lagrangian equations}

Using the above expressions for the brackets and inverting the 
system of equations (\ref{sistema}), we find the post--Newtonian 
Lagrangian planetary equations:
\begin{eqnarray}
&&\frac{d a}{dt} = \frac{2 c^2\,a^2}{G\,m}\, \left(1 + 
\frac{G\,m}{c^2\,a}\, \left(\frac{7}{2} - \frac{\nu}{2}\right)\right)\, 
\frac{\partial{\cal H}_1}{\partial T}
\label{L1}\\
\nonumber\\
&&\frac{d e}{dt} = \frac{c^2\,a (1 - e^2)}{G\,m e}\, \left(1 + 
\frac{G\,m}{c^2\,a}\, \left(\frac{11}{2} - \frac{3 \nu}{2}\right)\right)\, 
\frac{\partial{\cal H}_1}{\partial T} +\nonumber\\ 
&&+\frac{c^2\,\sqrt{1 - e^2}}{e \sqrt{G\,m a}}\, \left(1 + 
\frac{G\,m}{c^2\,a (1 - e^2)}\, \left(4 - \nu - e^2 + 
\frac{\nu e^2}{2}\right)\right)\, 
\frac{\partial{\cal H}_1}{\partial \omega}\\
\nonumber\\
&&\frac{d T}{dt} = - \frac{2 c^2\,a^2}{G\,m}\, \left(1 +
\frac{G\,m}{c^2\,a}\, \left(\frac{7}{2} - \frac{\nu}{2}\right)\right)\, 
\frac{\partial{\cal H}_1}{\partial a} - 
\frac{c^2\,a (1 - e^2)}{G\,m e}\, \left(1 +
\frac{G\,m}{c^2\,a}\, \left(\frac{11}{2} - \frac{3 \nu}{2}\right)\right)\, 
\frac{\partial{\cal H}_1}{\partial e}\nonumber\\ 
&&\\
&&\frac{d \omega}{dt} = - \frac{c^2\,\sqrt{1 - e^2}}{e \sqrt{G\,m a}}\, \left(1 
+ \frac{G\,m}{c^2\,a (1 - e^2)}\, \left(4 - \nu - e^2 +
\frac{\nu e^2}{2}\right)\right)\, 
\frac{\partial{\cal H}_1}{\partial e} +\nonumber\\
&&+ \frac{c^2\,\cot{i}}
{\sqrt{G\,ma} \sqrt{1 - e^2}}\, \left(1 - \frac{G\,m}{c^2\,a (1 - e ^2)}\, 
\left( 2 + e^2 - \frac{\nu e^2}{2}\right)\right)\, 
\frac{\partial{\cal H}_1}{\partial i}\\  
\nonumber\\
&&\frac{d \Omega}{dt} = - \frac{c^2}{\sqrt{G\,ma} \sqrt{1 - e^2} \sin{i}}\, 
\left(1 - \frac{G\,m}{c^2\,a (1 - e ^2)}\, \left( 2 + e^2 - 
\frac{\nu e^2}{2}\right)\right)\, \frac{\partial{\cal H}_1}
{\partial i}\\ 
\nonumber\\
&&\frac{d i}{dt} = - \frac{c^2\,\cot{i}}
{\sqrt{G\,ma} \sqrt{1 - e^2}}\, \left(1 - \frac{G\,m}{c^2\,a (1 - e ^2)}\, 
\left( 2 + e^2 - \frac{\nu e^2}{2}\right)\right)\, 
\frac{\partial{\cal H}_1}{\partial \omega} +\nonumber\\
&&+ \frac{c^2}{\sqrt{G\,ma} \sqrt{1 - e^2} \sin{i}}\, 
\left(1 - \frac{G\,m}{c^2\,a (1 - e ^2)}\, \left( 2 + e^2 - 
\frac{\nu e^2}{2}\right)\right)\, \frac{\partial{\cal H}_1}
{\partial \Omega}
\label{L6}
\end{eqnarray}
Once the external perturbation has been assigned, 
it is possible to evaluate its effects on the orbital parameters by
solving the system (\ref{L1})--(\ref{L6}); as in 
classical mechanics, the first--order time variation of orbital elements, 
is simply obtained after substitution of the unperturbed values into the 
right--hand sides of (\ref{L1})--(\ref{L6}). This could be done 
without assuming
that the magnitude of $\epsilon_{PN}$ is a first order term because the 
1PN effects have already been taken into account using the 
post--Keplerian parameterization instead of the Keplerian one.      

In celestial mechanics, the parameter $T$, is often replaced by
(e.g.~\cite{mac,browclem,roy}):
\begin{equation}
\sigma = -\,n\,T
\label{sigma}
\end{equation}
which is just the constant related to the {\em mean anomaly} 
$M = n t + \sigma$, i.e. the 
angle which the radius vector would have described if it had been 
moving uniformly with the average rate $n$.

This newly defined orbital element allows one to see that all 
Lagrangian brackets are not changed except of this one: 
\begin{equation}
\left[\sigma,\, a\right] = \frac{a}{2\,c^2}\, \sqrt{
\frac{G\,m}{a^3}}\, \left(1 + \frac{G\,m}{c^2\,a}\right)
\end{equation}
which takes the place of (\ref{aT}). Using this new set of 
elements the 1PN planetary equations become:
  
\begin{eqnarray}
&&\frac{d a}{dt} = -\frac{2 c^2\,a^2}{G\,m}\, 
\sqrt{\frac{G\,m}{a^3}}\, \left(1 - 
\frac{G\,m}{c^2\,a}\right)\, 
\frac{\partial{\cal H}_1}{\partial \sigma}
\label{dadt}\\
\nonumber\\
&&\frac{d e}{dt} = -\frac{c^2\,a (1 - e^2)}{G\,m e}\, 
\sqrt{\frac{G\,m}{a^3}}\, \left(1 + 
\frac{G\,m}{c^2\,a}\, \left(1 - \nu\right)\right)\, 
\frac{\partial{\cal H}_1}{\partial \sigma} +\nonumber\\ 
&&+\frac{c^2\,\sqrt{1 - e^2}}{e \sqrt{G\,m a}}\, \left(1 + 
\frac{G\,m}{c^2\,a (1 - e^2)}\, \left(4 - \nu - e^2 + 
\frac{\nu e^2}{2}\right)\right)\, 
\frac{\partial{\cal H}_1}{\partial \omega}\\
\nonumber\\
&&\frac{d \sigma}{dt} = \frac{2 c^2\,a^2}{G\,m}\, 
\sqrt{\frac{G\,m}{a^3}}\, \left(1 -
\frac{G\,m}{c^2\,a}\right)\, 
\frac{\partial{\cal H}_1}{\partial a} + 
\frac{c^2\,a (1 - e^2)}{G\,m e}\, 
\sqrt{\frac{G\,m}{a^3}}\, \left(1 +
\frac{G\,m}{c^2\,a}\, \left(1 - \nu\right)\right)\, 
\frac{\partial{\cal H}_1}{\partial e}\nonumber\\ 
&&\label{dsigmadt}\\
&&\frac{d \omega}{dt} = -\frac{c^2\,\sqrt{1 - e^2}}{e \sqrt{G\,m a}}\, \left(1 
+ \frac{G\,m}{c^2\,a (1 - e^2)}\, \left(4 - \nu - e^2 +
\frac{\nu e^2}{2}\right)\right)\, 
\frac{\partial{\cal H}_1}{\partial e} +\nonumber\\
&&+ \frac{c^2\,a \cot{i}}
{G\,m \sqrt{1 - e^2}}\, \sqrt{\frac{G\,m}{a^3}} 
\, \left(1 - \frac{G\,m}{c^2\,a (1 - e ^2)}\, 
\left( 2 + e^2 - \frac{\nu e^2}{2}\right)\right)\, 
\frac{\partial{\cal H}_1}{\partial i}\\  
\nonumber\\
&&\frac{d \Omega}{dt} = - \frac{c^2\,a}{G\,m \sqrt{1 - e^2} \sin{i}}\, 
\sqrt{\frac{G\,m}{a^3}}\, 
\left(1 - \frac{G\,m}{c^2\,a (1 - e ^2)}\, \left( 2 + e^2 - 
\frac{\nu e^2}{2}\right)\right)\, \frac{\partial{\cal H}_1}
{\partial i}\\ 
\nonumber\\
&&\frac{d i}{dt} = -\frac{c^2\,a \cot{i}}
{G\,m \sqrt{1 - e^2}}\, 
\sqrt{\frac{G\,m}{a^3}}\, \left(1 - \frac{G\,m}{c^2\,a (1 - e ^2)}\, 
\left( 2 + e^2 - \frac{\nu e^2}{2}\right)\right)\, 
\frac{\partial{\cal H}_1}{\partial \omega} +\nonumber\\
&&+ \frac{c^2\,a}{G\,m \sqrt{1 - e^2} \sin{i}}\, 
\sqrt{\frac{G\,m}{a^3}}\, 
\left(1 - \frac{G\,m}{c^2\,a (1 - e ^2)}\, \left(2 + e^2 - 
\frac{\nu e^2}{2}\right)\right)\, \frac{\partial{\cal H}_1}
{\partial \Omega}
\end{eqnarray}

Usually the disturbing function ${\cal H}_1$ depends upon  
$\sigma$ only through the phase $M~=n\,t + \sigma$,
therefore it is suitable to express ${\cal H}_1$ in terms of $M$. 
Furthermore ${\cal H}_1$ 
depends upon $a$ both explicitly and implicitly through $M$ 
(in fact $n = n(a)$), so that the derivative of the disturbing function 
taken with respect to $a$, can be written as follows:
\begin{equation}
\frac{\partial{\cal H}_1}{\partial a} = 
\left(\frac{\partial{\cal H}_1}{\partial a}\right)_M + 
\frac{\partial{\cal H}_1}{\partial M}\, t\, 
\frac{d n}{da};
\end{equation}
the first term on the right hand side is the derivative of 
${\cal H}_1$ with respect to $a$ while keeping $M$ (and then $n$)
as a constant, 
and the second term takes into account the dependence of ${\cal H}_1$
upon $a$ through $n$. 
From the above expression and using eq.~(\ref{dsigmadt}) 
we get: 
\begin{eqnarray}
&&\frac{d M}{dt} = n + \frac{d a}{dt}\, t\, \frac{d n}{da} + 
\frac{2 c^2\,a^2}{G\,m}\, \sqrt{\frac{G\,m}{a^3}}\, \left( 1 - 
\frac{G\,m}{c^2\,a}\right)\, 
\left(\frac{\partial{\cal H}_1}{\partial a}\right)_M +\nonumber\\ 
&&+ \frac{2 c^2\,a^2}{G\,m}\, \sqrt{\frac{G\,m}{a^3}}\, \left( 1 - 
\frac{G\,m}{c^2\,a}\right)\, \frac{\partial{\cal H}_1}{\partial M}\, 
t\, \frac{d n}{da} + \frac{c^2\,a (1 - e^2)}{G\,m e}\, 
\sqrt{\frac{G\,m}{a^3}}\, \left(1 + \frac{G\,m}{c^2\,a}\, \left( 1 - \nu\right)
\right)\, \frac{\partial{\cal H}_1}{\partial e};\nonumber\\
&&
\end{eqnarray}
Now, since
$\frac{\partial\ }{\partial M} = \frac{\partial\ }{\partial\sigma}$, 
eq.~(\ref{dadt}) becomes
\begin{equation}
\frac{d a}{dt} = -\frac{2 c^2\,a^2}{G\,m}\, 
\sqrt{\frac{G\,m}{a^3}}\, \left(1 - 
\frac{G\,m}{c^2\,a}\right)\, 
\frac{\partial{\cal H}_1}{\partial M}\\
\label{dadtnew}
\end{equation}
while the time derivative of $M$ can be written
\begin{eqnarray}
\frac{d M}{dt} &=& n + 
\frac{2 c^2\,a^2}{G\,m}\, \sqrt{\frac{G\,m}{a^3}}\, \left( 1 - 
\frac{G\,m}{c^2\,a}\right)\, 
\left(\frac{\partial{\cal H}_1}{\partial a}\right)_M 
+ \nonumber\\
&& + \frac{c^2\,a (1 - e^2)}{G\,m e}\, 
\sqrt{\frac{G\,m}{a^3}}\, \left(1 + \frac{G\,m}{c^2\,a}\, \left( 1 - \nu\right)
\right)\, \frac{\partial{\cal H}_1}{\partial e}
\label{dMdt}
\end{eqnarray}
Eqs.~(\ref{dadtnew}) and (\ref{dMdt}) take the 
place of (\ref{dadt}) and (\ref{dsigmadt}) when $M$ is used 
as orbital parameter (instead of $\sigma$ or $T$).

\section{Application}

As an example of application we consider the interaction of a
close binary system with a third body. For the sake of 
simplicity we assume the third body at a distance 
$R_B>>a$ from the binary centre of mass with mass $m_3>>m$. In 
this way the resulting motion could be safely approximated by the 
perturbed 1PN motion of the binary system around its centre of mass
$\bmit{r}_B$ which, in turns, has Keplerian orbit around $m_3$, 
considered motionless. The time evolution of $\bmit{r}_B$ is therefore 
known, this latter not being a variable of the problem. 
The discussion made at the beginning of section III showed that,  
in order
to achieve a consistent picture, the effect of the perturbation must 
be greater than that of the 2PN one.
The most interesting case is when the effect of the 
perturbation is of the same order of magnitude as the 1PN effect; as it 
will be seen later this occurs when
\begin{equation}
\frac{m_3}{R_B^3} \sim \frac{G\,m^2}{c^2\,a^4} \sim 
5 \left(\frac{M_\odot}{a.u.^3}\right)
\label{pert1pn}
\end{equation}
where the second relation holds true if $m \sim M_\odot$ and 
$a \sim 2\,R_\odot$.

Therefore, in the framework of previous assumptions, the perturbation to  
Hamiltonian~(\ref{H0}) is given by
\begin{equation}
{\cal H}_1 = - \frac{m_3\,G}{2\,c^2\,R_B^3} \left[
r^2 - 3 \frac{(\bmit{x}\cdot\bmit{R}_B)^2}{R_B^2}\right]
\label{example}
\end{equation}
where $\bmit{R}_B$ is the radius vector from $m_3$ to the 1PN centre 
of mass of the binary system.
We assume that the binary centre of mass performs a circular orbit 
around $m_3$,
which is coplanar with the 
orbital plane of the binary system. Hence we set
$i=0$, $\Omega=0$ and 
$\bmit{R}_B = A_B (\cos{\omega_3 t},\sin{\omega_3 t},0)$, where 
$\omega_3 = \sqrt{G\,m_3/A_B^3}$.
The Hamiltonian perturbation thus becomes
\begin{equation}
{\cal H}_1(a,e,M,\omega,t) =
-\,\frac{m_3\,G}{4\,c^2\,A_B^3}\,
a^2 (1-e \cos{\eta})^2 \left[
1 + 3 \cos{2\,(\omega_3\,t-\omega-\tilde\eta)}
\right]
\end{equation}
where $\eta = \eta(a,e,M)$ and 
$\tilde\eta = \tilde\eta(a,\eta(a,e,M),e)$ (see eqs.~(\ref{xtilde}) 
and~(\ref{Teta}), together with the definition of $M$ after 
eq.~(\ref{sigma})) and
\begin{equation}
\frac{\partial\eta}{\partial a} = 
\frac{(e-e_1)\,\sin{\eta}}{a\,(1-e_1 \cos{\eta})};\qquad
\frac{\partial\eta}{\partial e} = 
\frac{e_1 \sin{\eta}}{e\,(1-e_1 \cos{\eta})};\qquad
\frac{\partial\eta}{\partial M} = 
\frac{1}{1-e_1 \cos{\eta}}.
\label{detadaeM}
\end{equation}
From eqs.~(\ref{kappa}) and~(\ref{e3}) one also has
$\partial\kappa/\partial a = - \kappa/a$, 
$\partial\kappa/\partial e = 2\,e\,\kappa/(1-e^2)$,
$\partial e_3/\partial a = (e-e_3)/a$,
$\partial e_3/\partial e = e_3/e$. 
In order to obtain the equations providing the temporal derivatives of 
the orbital elements, 
it suffices to calculate the partial derivatives 
of ${\cal H}_1$; in this way the problem of motion is solved.
In order to to better understand the 
behaviour of the solution let us suppose $e=0$. In this case the 
Hamiltonian perturbation greatly simplifies
\begin{equation}
{\cal H}_1(a,M,\omega,t) =
-\,\frac{m_3\,G}{4\,c^2\,A_B^3}\,
a^2 \left[
1 + 3 \cos{2\,(\omega_3\,t-\omega-(1+\kappa) M)}
\right]
\end{equation}
Finally the equations for the orbital elements $a$ and $M$ are:
\begin{eqnarray}
\frac{da}{dt} &=& \frac{3\,a^2\,m_3}{A_B^3} \sqrt{\frac{G\,a}{m}}\,
(1+\kappa)\, \left( 1 -\frac{G\,m}{c^2\,a} \right) 
\sin{2\,(\omega_3\,t-\omega-(1+\kappa) M)}
\label{diadit}\\
\frac{dM}{dt} &=& n - \frac{m_3\,a}{A_B^3} \sqrt{\frac{G\,a}{m}}\,
\left( 1 -\frac{G\,m}{c^2\,a} \right) \left[
1 + 3\,\cos{2\,(\omega_3\,t-\omega-(1+\kappa) M)} \right]
\label{diMdit}
\end{eqnarray}
while
\begin{equation}
\frac{d(e^2)}{dt} = 0;\qquad
\frac{d\omega}{dt} = 0;\qquad
\frac{d\Omega}{dt} = 0;\qquad
\frac{di}{dt} = 0
\end{equation}
that is $e$, $\omega$, $\Omega$ and $i$ are constants of the 
motion.
By substituting the unperturbed values of the orbital elements in the 
right--hand side of eqs.~(\ref{diadit})--(\ref{diMdit}) the 
first--order time--variation of the parameters are achieved. 
As far as relation~(\ref{pert1pn}) is fulfilled, being 
$m_3\,a^3/(m\,A_B^3)\sim G\,m/(c^2\,a)$, the solution can be 
written as
\begin{eqnarray}
a(t) &=& a_0 \left(1 + 
\frac{3\,m_3\,a_0^3}{2\,m\,A_B^3}\,
\left[\cos{2\,(\omega_3\,t-\omega-(1+\kappa) M_0)} - 
\cos{2\,\omega} \right]\right)\label{sola}\\
M(t) &=& \left(1-\frac{m_3\,a_0^3}{m\,A_B^3}\right)\,M_0 +
\frac{3\,m_3\,a_0^3}{2\,m\,A_B^3}\,
\left[\sin{2\,(\omega_3\,t-\omega-(1+\kappa) M_0)} - 
\sin{2\,\omega} \right]\label{solM}
\end{eqnarray}
where $a_0$ and $M_0$ are the unperturbed values.

Accurate timing of the binary system could then provide the orbital 
profile through which (via eqs.~(\ref{sola})--(\ref{solM})) $m_3$ and 
$A_B$ are evaluated. 
This example has shown how the method proposed in this paper allowed 
one to achieve an already existing result in a relatively simple way.

\section{Conclusions}

We have developed a Post--Newtonian extension of the planetary 
Lagrangian equations in order to describe the motion of a relativistic 
binary system under the influence of a generic external perturbation.
An important class of phenomena that could be treated is the 
perturbation caused by external bodies to a binary system.
We also have provided an example featuring the capability of the 
method.

Our approach is a method to solve in a perturbative way the equations 
governing the variation of orbital parameters. The theory only assumes 
the smallness of the external perturbation while relativistic effects 
are already included in the unperturbed problem. This is the major 
advantage of our approach over classical Lagrangian methods.
In the present paper the problem is solved at 1PN level, but this 
approach should also be suitable 
to 2PN or higher--accuracy orders in the relativistic expansion.

\acknowledgments

The authors wish to thank V. Guidi for reading of the manuscript.

\end{document}